\documentclass[12pt]{article}

\usepackage{graphicx}

\advance \topmargin by -\headheight
\advance \topmargin by -\headsep
     
\evensidemargin \oddsidemargin
\def\nonu{\nonumber}

\def\br{\begin{eqnarray}}
\def\er{\end{eqnarray}} 
\def\be{\begin{equation}}
\def\bdm{\begin{displaymath}}
\def\ee{\end{equation}}
\def\edm{\end{displaymath}}
\def\0{\nonumber}
\def\lb{\lbrack}
\def\rb{\rbrack}
\def\({\left(}
\def\){\right)}
\def\v{\vert}

\relax

\def\a{\alpha}

\def\d{\delta}

\def\hc{\hat{c}}

\def\l{\lambda}

\def\o{\over}

\def\pa{\partial}
\def\pr{\prime}

\def\lie{{\cal G}}

\def\rlx{\relax\leavevmode}
\def\inbar{\vrule height1.5ex width.4pt depth0pt}
\def\IZ{\rlx\hbox{\sf Z\kern-.4em Z}}
\def\IR{\rlx\hbox{\rm I\kern-.18em R}}
\def\IC{\rlx\hbox{\,$\inbar\kern-.3em{\rm C}$}}
\def\one{\hbox{{1}\kern-.25em\hbox{l}}}

\def\NPB#1#2#3{{\sl Nucl. Phys.} {\bf B#1} (#2) #3}

\def\JMP#1#2#3{{\sl J. Math. Phys.} {\bf #1} (#2) #3}

\def\IJMPA#1#2#3{{\sl Int. J. Mod. Phys.} {\bf A#1} (#2) #3}

\def\PHSA#1#2#3{{\sl Physica} {\bf A#1} (#2) #3}

\def\JGP#1#2#3{{\sl J. of Geom. and Phys. } {\bf #1} (#2) #3}

\begin{document}

\begin{titlepage}
\vspace*{-2 cm}
\noindent
\begin{flushright}
\end{flushright}

\vskip 1 cm
\begin{center}
{\Large\bf Negative Even Grade mKdV Hierarchy and its Soliton Solutions } \vglue 1  true cm
{
{ J.F. Gomes}, G.Starvaggi Fran\c ca, G. R. de Melo and  { A.H. Zimerman}}\footnotetext{ jfg@ift.unesp.br, guisf@ift.unesp.br, gmelo@ift.unesp.br,  zimerman@ift.unesp.br}\\

\vspace{1 cm}

{\footnotesize Instituto de F\'\i sica Te\'orica - IFT/UNESP\\
Rua Dr. Bento Teobaldo Ferraz, 271, Bloco II\\
01140-070, S\~ao Paulo - SP, Brazil}\\

\vspace{1 cm}

\end{center}

\normalsize
\vskip 0.2cm

\begin{center}
{\large {\bf ABSTRACT}}\\
\end{center}
\noindent

In this paper we provide an algebraic  construction for the negative even mKdV 
 hierarchy which gives rise to  time evolutions associated to even graded Lie algebraic structure.
We propose a modification of  the dressing method, in order to incorporate a non-trivial 
vacuum configuration and construct a deformed  vertex
operator for $\hat{sl}(2)$, that enable us to obtain explicit and systematic solutions
for the whole negative even grade equations.

\noindent

\vglue 1 true cm

\end{titlepage}

\section{Introduction}
The odd mKdV hierarchy  consists of a series of non-linear equations 
of motion associated to certain odd graded Lie algebraic structure such that,
 each equation  correspond to a time evolution according to time 
$t=t_{2n+1}$ \cite{miwa}.
 Since these are directly associated to  odd graded  operators,
  they are dubbed {\it odd order mKdV hierarchy}.
 
 In this paper we employ  the algebraic  technique which, 
 for positive order, the graded structure of the zero curvature 
 representation imposes severe restrictions so that only odd times 
  are allowed.  For negative order however, the structure is less 
restrictive and gives also rise to a subclass of equations of motion, 
described by   time evolutions associated to {\it negative even grades}.
 These are  constructed by  the  Lax operator in terms of an  affine graded Lie algebra, $\hat {sl}(2)$
which, from  the  zero 
curvature representation  generates  systematically   a series of nonlinear integrable equations.

By considering a special case of Zakharov-Shabat AKNS spectral problem and 
using recursion techniques, a class of integrable equations were 
considered \cite{qiao} in order to develop negative order mKdV hierarchy as 
well as to  obtain  some parametric type of solutions.

Here, in our approach, the simplest case of the even mKdV where 
$t = t_{-2}$, is studied in detail and a crucial observation that a 
trivial zero constant 
solution is not admissible, lead us to extend the dressing method to 
incorporate non-zero constant vacuum solutions.
This implies in deforming the usual vertex operators  preserving 
the nilpotency property peculiar in solving for soliton solutions.  
Employing the modified dressing formalism, we construct multi-soliton
solutions for the whole negative even grade mKdV hierarchy.


In Sect. 2 we discuss the algebraic formalism  for positive  and negative 
hierarchies \cite{ frank, modugno}. In particular, the  construction of the 
equation of motion 
for $t=t_{-2}$. Such equation agrees with the one proposed in 
\cite{qiao} using recursion operator techniques. In Sect. 3  we discuss 
the dressing formalism \cite{nissimov, mira, babelon, olive, ferreira} to 
construct soliton 
solutions for the odd hierarchy. In this case the formalism is based  
upon a constant zero vacuum solution  which, by   gauge transformation,  
generate multi-soliton solutions.  In Sect. 4 we  extend the dressing 
formalism  to the negative even hierarchy, by introducing a non-zero  
constant vacuum configuration.
We then construct the deformed vertex operators which generate explicitly 
the multi-soliton solutions. Some details involving the explicit  calculation  of 
matrix elements  of products  of   vertex operators  and the proof of their nilpotency are 
described in the appendix.

\section{Positive and Negative Hierarchies}

Consider the {\it positive   mKdV hierarchy}  given by the zero curvature 
representation
\be
\lb \pa_x + E^{(1)} + A_0, \pa_{t_n} + D^{(n)} + D^{(n-1)} + 
\cdots +D^{(0)} \rb = 0,
\label{1}
\ee
where $E^{(2n+1)} = \l^n \(E_{\a} + \l E_{-\a}\), \; A_0 = v h$  
contains the field  variable $v=v\(x,t_n\)$, 
and $\{E_{\pm \a}, h\}$ are $sl(2)$ generators 
satisfying $\lb h, E_{\pm \a} \rb = \pm 2E_{\pm \a}, \;
\lb E_{\a}, E_{-\a} \rb= h$.  
The grading operator $Q = 2\l{{d} \o {d\l}} + {1 \o 2} h$ decomposes  the 
affine Lie algebra $\hat{sl}(2)$ into graded subspaces, 
$\hat\lie = \oplus_{i}\lie_i$,  
\be
\lie_{2m} = \{ h^{(m)} = \l^m h\}, \qquad 
\lie_{2m+1} = \{\l^m\(E_{\a} + \l E_{-\a}\), \; \l^m\(E_{\a} - \l E_{-\a}\)\}
\label{2}
\ee
$m=0,\pm1,\pm2,\ldots$ and  $D^{(j)} \in \lie_j$.  A more subtle structure arises 
if one consider the decomposition  $\hat\lie = {\cal K} \oplus {\cal M}$ 
where ${\cal K} =\{ \l^n\(E_{\a} + \l E_{-\a}\)\} $ denotes the Kernel of 
$E \equiv E^{(1)}$, i.e., 
${\cal K} = \{ k \in {\hat \lie} \; \v \; \lb E, k\rb =0\}$ and
${\cal M}$ is its complement.  
We assume that $E$ is semi-simple in the sense that this second decomposition 
is such that 
\be
\lb  {\cal K} , {\cal K} \rb \subset  {\cal K}, \qquad
\lb  {\cal K} , {\cal M} \rb \subset  {\cal M}, \qquad
\lb  {\cal M} , {\cal M} \rb \subset  {\cal K}.
\label{3}
\ee
Eqn. (\ref{1}) can  be decomposed grade by grade and solved for $D^{(j)}$.
For instance, the highest grade in (\ref{1}) yields
\be
\lb E, D^{(n)} \rb = 0 \Rightarrow D^{(n)} = D^{(n)}_{\cal K} \in {\cal K}.
\label{4}
\ee
Since by (\ref{2}) $\cal K$ has grade $2m+1$, this last equation implies 
that $n=2m+1$ and hence $t_n = t_{2m+1}$, showing that only {\it odd} grades 
are admissible for positive mKdV hierarchy (\ref{1}). 
Moving down grade by grade and 
using the symmetric space structure (\ref{3}), eqn. (\ref{1}) allows one to 
solve for all $D^{(j)} = D_{\cal K}^{(j)} + D_{\cal M}^{(j)}, \; j=0 \ldots n$.
In particular, the zero grade projection in ${\cal M}$ yields the equation of 
motion
\be
\pa_{t_{n}}A_0-\pa_x D_{\cal M}^{(0)} - \lb A_0, D_{\cal K}^{(0)}\rb = 0
\label{5}
\ee
where we have taken into account that $A_0 \in {\cal M}$.
Eqn. (\ref{5}) represents a series of nonlinear evolution equations 
associated with time $t_{2m+1}$. Choosing $m=1$ for example, we will obtain
the well known mKdV equation \cite{miwa, frank}.

The same, however, does no happen for the {\it negative mKdV hierarchy}
\cite{frank, modugno}, i.e., for  $n<0$. Let us consider the zero curvature 
representation
\be
\lb \pa_x + E^{(1)} + A_0, \pa_{t_{-n}} + D^{(-n)} + D^{(-n+1)} 
+\cdots +D^{(-1)} \rb = 0.
\label{6}
\ee
Here, the lowest grade projection,
\be
\pa_x D^{(-n)} + \lb A_0, D^{(-n)}\rb = 0
\label{7}
\ee
yields a nonlocal equation for $D^{(-n)}$. The second lowest projection 
of grade $-n+1$ leads to 
\be
\pa_x D^{(-n+1)} + \lb A_0, D^{(-n+1)} \rb + \lb E^{(1)},D^{(-n)} \rb = 0
\label{8}
\ee
which determines $D^{(-n+1)}$. The same mechanism works recursively until we reach the zero 
grade equation 
\be
\pa_{t_{-n}} A_0  + \lb  E^{(1)},D^{(-1)}\rb  = 0
\label{9}
\ee
which  gives the time evolution for the field in $A_0$ according to 
time $t_{-n}$.  
The simplest model of this sub-hierarchy is obtained for $n=1$,  
for which the following equations arrive from the zero curvature  (\ref{6}),
\br
\pa_x D^{(-1)} + \lb A_0, D^{(-1)}\rb & = & 0, \nonu \\
\pa_{t_{-1}} A_0  - \lb  E^{(1)},D^{(-1)}\rb  & = & 0.
\label{10}
\er
These equations can be solved in general if we parametrize the fields as
\be
D^{(-1)} = B^{-1} E^{(-1)} B, \qquad A_0 = B^{-1}\pa_x B, 
\qquad B = \exp (\lie_0)
\label{11}
\ee
in terms of the zero grade subalgebra $\lie_0$. Space-time is associated 
to the light-cone coordinates $\bar z, z$ as $x = \bar z, \; 
t_{-1} = z$.
The time evolution is  then given by the Leznov-Saveliev equation, 
\be
\pa_{t_{-1}} \(B^{-1}\pa_x B\) = \lb  E^{(1)}, B^{-1} E^{(-1)}B \rb 
\label{12}
\ee
which for $\hat{sl}(2)$ with principal gradation 
$Q= 2\l {{d}\o {d\l}} + {1\o 2} h$, yields the sinh-Gordon equation 
\cite{babelon}
\be
\pa_{t_{-1}} \pa_x \phi  = e^{2\phi} - e^{-2\phi}, \qquad  B= e^{\phi h}.
\label{13}
\ee
Note that from the definition of $A_0$ and the parametrization (\ref{11}), 
we find the following relation between $\phi$ and $v$: 
$A_0=vh=B^{-1}\partial_x B\; \Rightarrow \; v=\partial_x \phi$.

We now propose the first nontrivial example for the {\it negative even} 
sub-hierarchy:
\br
\pa_x D^{(-2)} + \lb A_0, D^{(-2)}\rb & = & 0, \label{14.a} \\
\pa_x D^{(-1)} + \lb A_0, D^{(-1)}\rb + \lb  E^{(1)},D^{(-2)}\rb &=&0, 
\label{14.b} \\
\pa_{t_{-2}} A_0  - \lb  E^{(1)},D^{(-1)}\rb & = & 0. \label{14.c}
\er
From grading (\ref{2}) and decomposition (\ref{3}) we find 
\br
D^{(-2)} &=& c_{-2} \l^{-1} h,\nonu\\
D^{(-1)} &=& a_{-1} \(\l^{-1} E_{\a} + E_{-\a}\) + b_{-1} \(\l^{-1} E_{\a} - 
E_{-\a}\). 
\label{15}
\er
From eqn. (\ref{14.a}) and  (\ref{14.b})  we find,  $c_{-2}=\textnormal{const.}$ and 
\br
\pa_x \( a_{-1} + b_{-1}\) + 2 v \( a_{-1} + b_{-1}\) -2c_{-2}&=& 0, \nonu \\ 
\pa_x\( a_{-1} - b_{-1}\)  -2v \(a_{-1}- b_{-1}\) +2 c_{-2} &=&0,
\label{16}
\er
which are ordinary differential equations with solution
\br
a_{-1} + b_{-1} &=& 
    2c_{-2} \exp ( -2 d^{-1} v) d^{-1} \( \exp (2 d^{-1} v) \), \nonu \\
a_{-1} - b_{-1} &=& 
    -2c_{-2} \exp ( 2 d^{-1} v) d^{-1} \( \exp (-2 d^{-1} v) \).
\label{17}
\er
In (\ref{17})  we have denoted $d^{-1} f = \int^x f(x^{\pr} )dx^{\pr}$.  
Having determined  $D^{(-1)} $, the evolution equation associated to 
time $t_{-2}$ is then given by eqn. (\ref{14.c}):
\be
\pa_{t_{-2}} v + 2c_{-2} e^{- 2d^{-1} v } d^{-1} \( e^{2 d^{-1} v} \)+  
2c_{-2} e^{2d^{-1} v } d^{-1} \( e^{-2 d^{-1} v}\) = 0.
\label{18}
\ee
Differentiating twice with respect to $x$, and setting $c_{-2}=1$ for 
convenience, we find the local equation 
\be
 v_{xx t_{-2}} -4 v^2 v_{t_{-2}} -  {{v_x v_{x t_{-2}}}\o {v}} 
 - 4 {{v_x}\o {v}} = 0.
\label{19}
\ee
Eqn. (\ref{19}) was already obtained in \cite{qiao} using recursion operator techniques.

\section{Odd Hierarchy Solutions}

In order to  employ the dressing method to construct soliton 
solutions, we now introduce the full $ \hat {sl}(2)$  affine Kac-Moody algebra  
with central extensions:
\br
\lb h^{(m)}, h^{(n)} \rb &=& 2 m\d_{m+n,0}\hat{c}, \nonu \\
\lb h^{(m)}, E_{\pm \a}^{(n)} \rb &=& \pm 2E_{\pm \a}^{(m+n)}, \nonu \\
\lb E_{\a}^{(m)}, E_{-\a}^{(n)} \rb&=& h^{(m+n)} + m\d_{m+n,0}\hat{c}
\label{19a}
\er 
together with the derivation operator $\hat d$ such that,
\be
\lb \hat d, T_a^{(n)} \rb = n T_a^{(n)}, \qquad 
T_a^{(n)} = \{h^{(n)}, \;\; E_{\pm \a}^{(n)}\}.
\ee
The grading operator now reads $Q = 2\hat d + 1/2 h^{(0)}$.
A well established method for determining soliton solutions is 
 to choose a vacuum solution and then to  map it into  
a non trivial solution by gauge transformation (dressing) 
\cite{mira},\cite{babelon}.  
The zero curvature condition (\ref{1}) or (\ref{6}) implies pure gauge 
connections, $A_x = T^{-1}\pa_x T = E + A_0 $ and 
$ A_{t_n} = T^{-1} \pa_{t_n}  T  = D^{(n)} + \cdots + D^{(0)}$ or 
$ A_{t_{-n}} =  T^{-1}\pa_{t_{-n}} T = D^{(-n)} + \cdots + D^{(-1)}$, 
respectively.  Suppose there exists a vacuum solution satisfying 
\be
{A}_{x, vac}=E^{(1)} -  t_k \d_{k+1,0}\hat{c}, \qquad
A_{t_k, vac}=E^{(k)},  
 \label{8.15}
\ee
where now $[E^{(k)},E^{(l)}]={{1}\o {2}}(k-l) \d_{k+l,0}\hat{c}$, for $(k,l)$ 
odd integers.
 
The solution for 
$ A_{x, vac} =  T_0^{-1}\pa_x T_0 $ and
$A_{t_{k}, vac} = T^{-1}_0\pa_{t_k} T_0 $  
is therefore given by 
\be
T_{0} =\exp (xE^{(1)}) \exp (t_k E ^{(k)}).
\label{8.18}
\ee
The dressing method is based on the assumption of the existence of two gauge
transformations, generated by $\Theta_{\pm}$, mapping the vacuum into non 
trivial configuration, i.e.
\br
{A}_x & = & ({\Theta_{\pm }})^{-1}{A}_{x, vac}\Theta_{\pm }+ 
({\Theta_{\pm }})^{-1}{\partial_x }\Theta_{\pm },  \label{8.20} \\
A_{t_k} & = & ({\Theta_{\pm }})^{-1}A_{t_k, vac}\Theta_{\pm }+
({\Theta_{\pm }})^{-1}\partial_{t_k} \Theta_{\pm }. \label{8.19} 
\er
As a  consequence we relate
\be
\Theta_{-}\Theta_{+}^{-1} = T_0^{-1}gT_0  \label{8.24}
\ee
where $g$ is an arbitrary constant group element.
We suppose  that $\Theta_{\pm}$ are group elements of the form
\be
\Theta_{-}^{-1}=e^{p(-1)}e^{p(-2)} \ldots\: , \qquad 
\Theta_{+}^{-1}=e^{q(0)}e^{q(1)}e^{q(2)} \ldots
\label{8.26}
\ee
where $p^{(-i)}$ and $q^{(i)}$ are linear combinations of grade 
$(-i)$ and $(i)$ generators, respectively ($i=0,1,\ldots$).
In considering $\Theta_{+}$, the zero grade component of (\ref{8.20})
admits solution
\be
e^{q(0)} = B^{-1}e^{-\nu\hat c}
\label{t0}
\ee
where we have used $A_x = E^{(1)} + B^{-1}\pa_x B + \pa_x\nu\hat{c}
-t_k\delta_{k+1,0}\hat{c}$.
From eqn. (\ref{8.24}) we find 
\be
\ldots e^{-p(-2)}e^{-p(-1)}B^{-1}e^{-\nu\hat{c}}e^{q(1)}e^{q(2)} \ldots = 
T_{0}^{-1}gT_{0}  \label{8.43}
\ee
hence, 
\be
<\lambda^{\pr} |B^{-1}|\lambda >e^{-\nu} =\;
< \lambda^{\pr}|T_{0}^{-1}gT_{0}|\lambda >  \label{8.45}
\ee
where $| \l>$ and $<\lambda^{\pr}|$ are annihilated by  $\lie_>$ and 
$\lie_<$, respectively.
Explicit space time dependence for the field in $\lie_0$, defined in 
(\ref{13}), is given by choosing specific matrix elements 
(\ref{highest_states}):
\br
e^{-\nu} &=& < \lambda_0|T_{0}^{-1}gT_{0}|\lambda_0 >,
\nonu \\
e^{-\phi - \nu} &=& < \lambda_1|T_{0}^{-1}gT_{0}|\lambda_1 >.
\label{tau}
\er
where $|\l_i>, i=0,1$ correspond to highest  weight states, i.e. annihilated by positive grade operators.
Suppose we now write the constant group element $g$ as 
\be
g=\exp\{F\(\gamma\)\},  \label{9.1}
\ee
where $\gamma$ is a complex parameter and we choose $ F\(\gamma \)$ to be an 
eigenstate of $E^{(k)}$, i.e.         
\be
\lbrack E ^{(k) },F\(\gamma \)]=f^{(k)}\(\gamma \)F\(\gamma \)  \label{9.2}
\ee
where $f^{(k)}$ are specific functions of $\gamma$. It therefore follows that
\be
T_{0}^{-1}gT_{0}=\exp\{\rho \(\gamma \)F\(\gamma \)\}  \label{9.6}
\ee
where 
\be
\rho\(\gamma\) =\exp\{-t_kf^{(k)}\(\gamma\)-xf^{(1)}\(\gamma\)\}.  \label{9.5}
\ee
For more general cases in which
\be
g=\exp\{F_{1}\(\gamma _{1}\)\}\exp\{F_{2}\(\gamma _{2}\)\}\ldots
\exp\{F_{N}\(\gamma_{N}\)\}  \label{9.7}
\ee
with
\be
\lb E ^{(k)},F_{i}\(\gamma _{i}\)\rb = f_i^{(k) }\(\gamma_i\)F_i\(\gamma_i\)
\label{9.8}
\ee
we find
\be
T_0^{-1}gT_0=
\exp\{\rho_1\(\gamma_1\)F_1\(\gamma_1\)\}
\exp\{\rho_2\(\gamma_2\)F_2\(\gamma_2\)\}\ldots
\exp\{\rho_N\(\gamma_N\)F_N\(\gamma_N\)\}
\label{9.9}
\ee
where 
\be
\rho_i\(\gamma_i\)=\exp\{-t_k f_i^{(k)}\(\gamma_i\)
-xf_i^{(1)}\(\gamma_i\)\}.  \label{9.10}
\ee
The specific eigenstate, in this case of $\hat{sl}(2)$, is given by
\be
F\(\gamma\)=\sum_{n=-\infty}^{\infty} \(h^{(n)} - 
{1\o 2} \d_{n,0}\hat{c}\) \gamma^{-2n} + 
\( E_{\a}^{(n)} - E_{-\a}^{(n+1)}\) \gamma^{-2n-1}
\label{vertice}
\ee
whose eigenvalues are obtained from
\be
\lb E ^{(k)},F\(\gamma\) \rb = -2\gamma ^{k}
F\(\gamma\).  \label{9.72}
\ee
From eqns. (\ref{tau}) we obtain solutions for $\phi$ (or equivalently 
$v=\pa_x\phi$) for the whole odd hierarchy, i.e, 
for all variables $t_k$ in eq. (\ref{9.10}).
Observe that in eqns. (\ref{9.2}) and (\ref{9.8}), $F\(\gamma\)$ is a
simultaneous eigenstate of both $E^{(1)}$ and $E^{(k)}$  and belong to the kernel  ${\cal K}$.
 The above argument is therefore valid only for 
$k=2m+1, \; m=0,\pm 1, \pm 2, \ldots$
since $ {\cal K}$ contains only odd grade elements. Then, this method
gives explicit solutions for both, eqns. (\ref{1}) and (\ref{6}) 
for  $n=k=2m+1$. See for instance \cite{babelon, olive, ferreira}.

\section{Negative Even Hierarchy Solutions}
 
In order to modify the dressing method to construct systematic solutions of 
equations like (\ref{18}) or (\ref{19}), we notice that $v=0$ cannot be 
solution of (\ref{19}).  Therefore,
let us propose the simplest vacuum configuration 
\br
{A}_{x, vac} & = & \(E^{(0)}_{\a} + E^{(1)}_{-\a}\) + v_0 h^{(0)} - 
\frac{1}{v_0}t_{-2m}\delta_{m-1,0}\hat{c}, \nonu \\
A_{t_{-2m}, vac} & = & 
\frac{1}{v_0}\(E^{(-m)}_{\a} +E^{(1-m)}_{-\a}\) + h^{(-m)}
\label{n1}
\er
with $v_0=\textnormal{const.}\neq 0$. It is straightforward to verify 
the zero curvature equation
\be
\lb\partial_x + A_{x, vac}, \partial_{t_{-2m}} + A_{t_{-2m}, vac}\rb=0.
\ee

This \emph{nontrivial  vacuum} leads to the following 
modification of eqn. (\ref{8.18}), but now for negative even grades,
\be
T_{0} = \exp\left\{x\(E^{(0)}_{\a} + E^{(1)}_{-\a}+v_0h^{(0)}\)\right\} 
\exp\left\{\frac{t_{-2m}}{v_0}\(E^{(-m)}_{\a} + 
E^{(1-m)}_{-\a}+v_0h^{(-1)}\)\right\}.
\label{n2}
\ee
The analogous of eqn. (\ref{t0}) leads  to 
\br
e^{q(0)} = B^{-1} e^{xv_0h^{(0)}} e^{-\nu\hat{c}}.
\label{n3}
\er
Observe that consistency of the zero curvature representation with 
nontrivial vacuum configuration requires terms with mixed gradation in 
constructing $T_0$ as in (\ref{n2}). The solution is then given by 
\br
e^{-\nu} &=& < \lambda_0|T_{0}^{-1}gT_{0}|\lambda_0 > \;\; \equiv \;\;
\tau^+,\nonu \\
e^{-\phi + xv_0 - \nu} &=& < \lambda_1|T_{0}^{-1}gT_{0}|\lambda_1 > 
\;\; \equiv \;\; \tau^-
\label{tau0}
\er
and hence,
\be
v = v_0 - \pa_x \ln\( \frac{\tau^+}{\tau^-} \), \qquad v=\partial_x\phi.
\label{sol}
\ee

In order to construct explicit soliton solutions we need a
simultaneous eigenstate of 
$b_1 \equiv E^{(0)}_{\a} + E^{(1)}_{-\a}+v_0h^{(0)}$ and 
$b_{-2m} \equiv \(E^{(-m)}_{\a} + E^{(-m+1)}_{-\a}+v_0h^{(-m)}\)$.
Let  
\be
F\(\gamma, v_0\) = \sum_{n=-\infty}^{\infty}\(\gamma^2-v_0^2\)^{-n}
\bigg[ h^{(n)}+\frac{v_0-\gamma}{2\gamma}\delta_{n,0}\hc
+E_{\a}^{(n)}\(\gamma+v_0\)^{-1}-E_{-\a}^{(n+1)}\(\gamma-v_0\)^{-1}\bigg].
\label{vertex_v0} 
\ee
be our deformed  vertex operator.
A direct calculation shows that
\br
\lb b_1, F\(\gamma, v_0\) \rb & = & -2\gamma F\(\gamma, v_0\), \nonu \\
\lb b_{-2m}, F\(\gamma, v_0\) \rb & = & -2\gamma\(\gamma^2-v_0^2\)^{-m} 
F\(\gamma, v_0\).
\label{eigenvalue_v0}
\er
Therefore from (\ref{9.5})  we find,
\be
\rho\(\gamma,v_0\) = \exp\left\{
2\gamma x + \frac{2\gamma t_{-2m}}{v_0\(\gamma^2-v_0^2\)^{m}}
\right\}.
\ee

It only remains to calculate the matrix
elements in eqns. (\ref{tau0}). They are shown in Appendix.
Note that, because of the nilpotency property of the vertex operator
between matrix elements, as discussed in the Appendix, the exponential
series in eqn. (\ref{tau0}) truncates, e.g, if we take,
\be
g = \exp\{F(\gamma, v_0)\}
\label{one_vertex}
\ee
we have
\br
<\lambda_a|T_0^{-1}gT_0|\lambda_a> &=& <\lambda_a|
\exp\left\{ \rho\(\gamma, v_0\) F\(\gamma, v_0\) \right\} 
|\lambda_a> \nonu \\
&=& 1 + \rho\(\gamma, v_0\)<\lambda_a| F\(\gamma, v_0\) |\lambda_a>.
\er
Thus, from eqn. (\ref{sol}), we obtain {\it explicit} solutions for the 
{\it whole negative even hierachy}.
The introduction of a nontrivial vacuum configuration, $v_0$, in
the dressing method, seems to have the same effect as a change in the
boundary conditions when looking for solutions of differential equations, 
as can be noted in eqn. (\ref{sol}). In this vein, we can say that our 
modified dressing approach implements a different boundary condition than 
the usual dressing of a trivial vacuum, largely used until now.

Let us introduce the shorthand notation:
\br
c^{\pm}_i & = & \frac{v_0\pm\gamma_i}{2\gamma_i}, \nonu \\
a_{ij} & = & 
\(\frac{\gamma_i-\gamma_j}{\gamma_i+\gamma_j}\)^2, \nonu \\
\rho_i & = & \exp\left\{ 2\gamma_i x + 
\frac{2\gamma_i t_{-2m}}{v_0\(\gamma_i^2-v_0^2\)^{m}} \right\}.
\label{data}
\er
The 1-soliton solution in eqn. (\ref{sol}), is obtained with one
vertex as in (\ref{one_vertex}). The explicit tau functions are:
\be
\tau^{\pm} = 1 + c_1^{\pm}\rho_1.
\label{1soliton}
\ee
The 2-soliton solution is obtained with
\be
g = \exp\left\{F\(\gamma_1, v_0\)\right\}\exp\left\{F\(\gamma_2, v_0\)\right\}
\ee
in (\ref{tau0}), and then
\be
\tau^{\pm} = 1 + c_1^{\pm}\rho_1 + c_2^{\pm}\rho_2 + 
c_1^{\pm}c_2^{\pm}a_{12}\rho_1\rho_2.
\label{2soliton}
\ee
The 3-soliton solution, obtained as a product of 3 exponential vertices, 
is given by
\br
\tau^{\pm} & = & 1 + c_1^{\pm}\rho_1 + c_2^{\pm}\rho_2 + c_3^{\pm}\rho_3 +
\nonu \\ 
& + & c_1^{\pm}c_2^{\pm}a_{12}\rho_1\rho_2 + 
c_1^{\pm}c_3^{\pm}a_{13}\rho_1\rho_3 + 
c_2^{\pm}c_3^{\pm}a_{23}\rho_2\rho_3 + \nonu \\
& + & c_1^{\pm}c_2^{\pm}c_3^{\pm}a_{12}a_{13}a_{23}\rho_1\rho_2\rho_3.
\label{3soliton}
\er

If we then substitute
\be
g = \prod_{i=1}^{n}\exp\{F\(\gamma_i, v_0\)\}
\ee
in eqns. (\ref{tau0}) we obtain the general n-soliton solution:
\be
\tau^{\pm} = \sum_{J \subset I}\(\prod_{i\in J}c_i^{\pm}\)
\(\prod_{ i,j \in J, \: i<j }
a_{ij}\)\prod_{i\in J}\rho_i
\label{Nsoliton}
\ee
where $I=\{1,\ldots,n\}$ and the sum is over all subsets $J$ of $I$. These
solutions present the same structure as those constructed from the trivial 
vacuum solution.  They differ only  by the deformation in  (\ref{data}) which now incorporates  the  parameter $v_0$.

\begin{figure}
\centering
\includegraphics[scale=0.86]{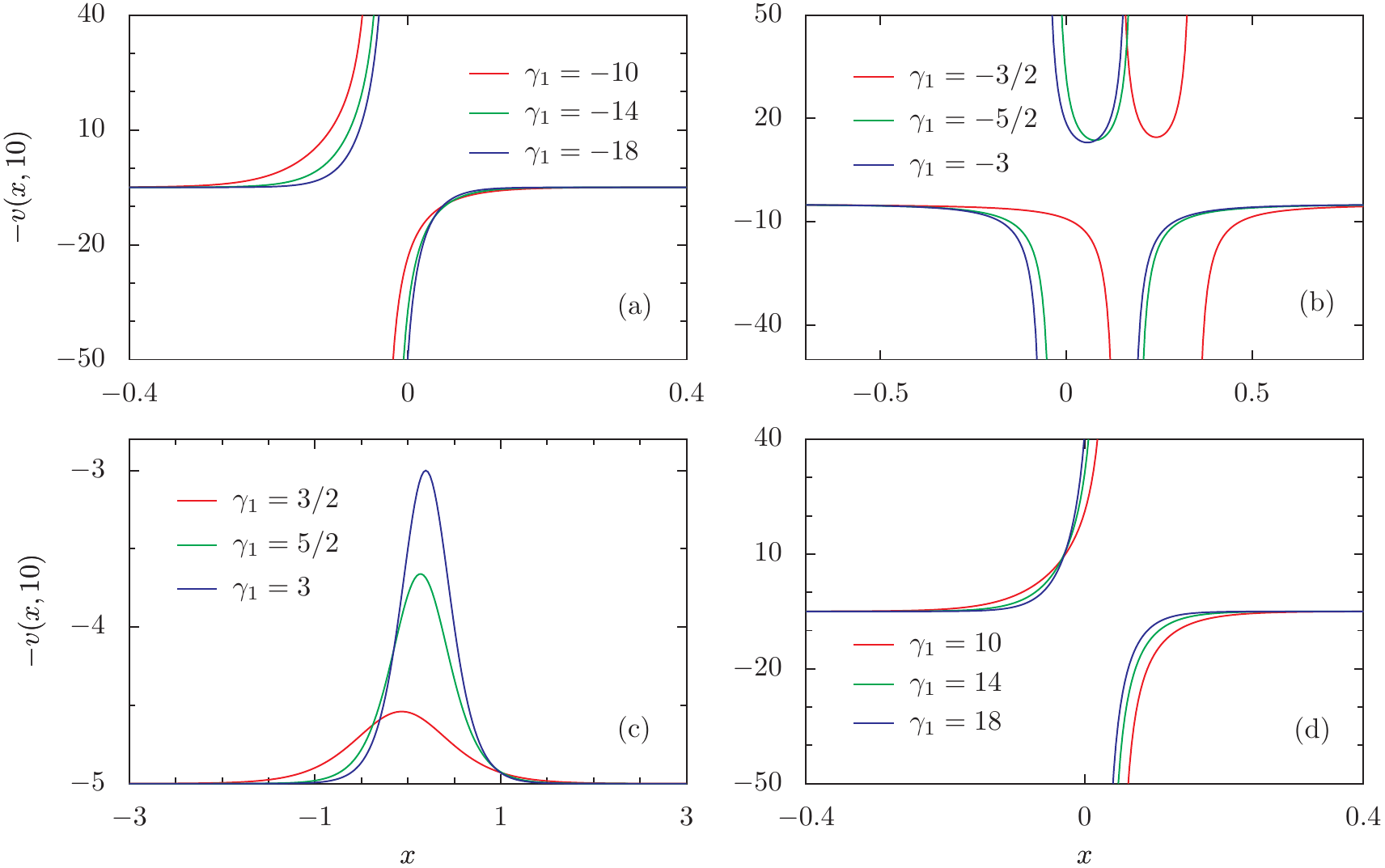}
\caption{
1-soliton solutions for eqn. (\ref{19}), $t_{n}=t_{-2}$ in mKdV hierarchy.
In all these graphs we set $v_0=5$ and choose a fixed time $t_{-2}=10$.
}
\label{fig:1soliton}
\end{figure}

The solutions of eqn. (\ref{19}), $t_{-2m}=t_{-2}$, are obtained by
setting  $m=1$ in (\ref{data}). Considering
1-soliton solution (\ref{1soliton}), we see that a critical behavior 
occurs when $\gamma_1\to \pm v_0$ or $\gamma_1\to 0$. 
So, we have $4$ different regions to consider: 
$\gamma_1 < -v_0$; $-v_0 < \gamma_1 < 0$; $0 < \gamma_1 < v_0$;
$\gamma_1 > v_0$. All these regions are considered separately in
Fig. (\ref{fig:1soliton}). These solutions keep their
form for any $t_{-2}$. Note that 3 different types of behavior occur, 
Fig. (\ref{fig:1soliton}-a) and Fig. (\ref{fig:1soliton}-d) are of the
same type, and have the same form as the usual \emph{trivial vacuum} 
solutions of \emph{odd grade} mKdV hierarchy, except from the fact
that the solution is displaced by  $v_0$ in the 
$y-\textnormal{axis}$. Fig. (\ref{fig:1soliton}-b)
and Fig. (\ref{fig:1soliton}-c) are different ones, and their form are
not obtained from the trivial vacuum configuration. Also, note
that $v \to v_0$ when $x\to \pm \infty$.

\begin{figure}
\centering
\includegraphics[scale=0.86]{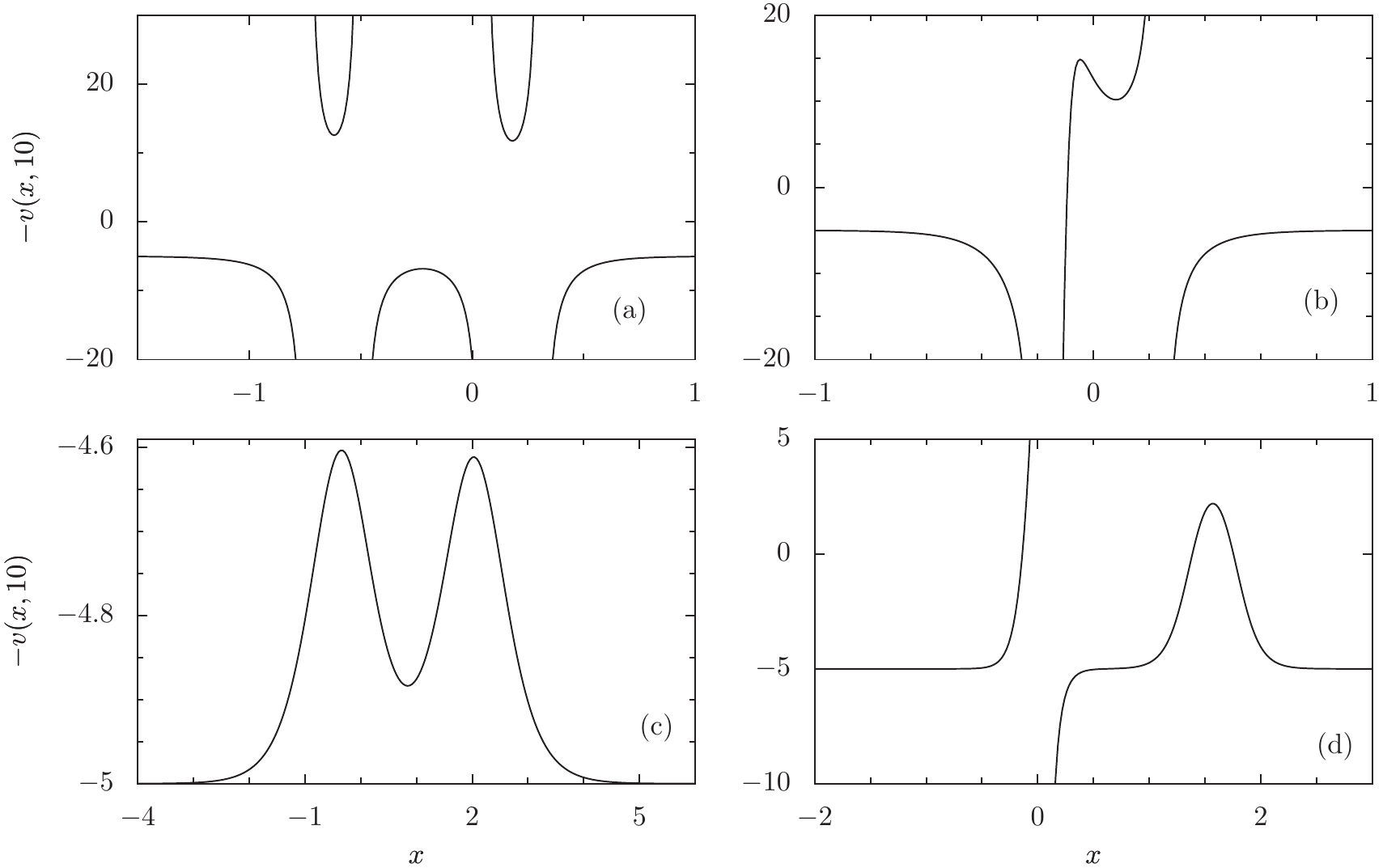}
\caption{2-soliton solutions for eqn. (\ref{19}).
$v_0=5$ and $t_{-2}=10$. Parameters: 
(a) $\gamma_1=-4, \; \gamma_2=-3$;
(b) $\gamma_1=-4, \; \gamma_2=-10$;
(c) $\gamma_1=1.3, \; \gamma_2=1.5$;
(d) $\gamma_1=4.8, \; \gamma_2=7$.}
\label{fig:2soliton}
\end{figure}

In Fig. (\ref{fig:2soliton}) we show the 2-soliton solution 
for eqn. (\ref{19}), where we illustrate the mixing of different type 
of solutions. Again, a different behavior emerges compared with
the trivial vacuum 2-soliton solutions.

\section{Conclusions}
We extended the mKdV hierarchy to include \emph{negative even grade} equations,
based on a graded infinite dimensional Lie algebra $\hat {sl}(2)$. This procedure systematically lead us
to obtain new non-linear integrable equations, e.g. eqn. (\ref{19}) which  was previously  obtained 
in \cite{qiao}. Our method  can also provide  other higher order integro-diffential  equations, 
like for example:
\br
\partial_x\partial_{t_{-4}}\phi & = & 
4 e^{-2\phi}d^{-1}\left[ e^{2\phi}d^{-1}\(e^{-2\phi}d^{-1}e^{2\phi} + 
e^{2\phi}d^{-1}e^{-2\phi}\) \right] - \nonu \\
&-& 4 e^{2\phi}d^{-1}\left[ e^{-2\phi}d^{-1}\(e^{-2\phi}d^{-1}e^{2\phi} + 
e^{2\phi}d^{-1}e^{-2\phi}\) \right].
\label{mkdvt-4}
\er

This  subhierarchy
of even grade equations are not solved by the usual dressing method, based
on a trivial vacuum configuration. Nevertheless, we also extended the 
dressing method to incorporate a constant non trivial vacuum configuration
$v_0$.

Remarkably, all these modifications lead us to obtain solutions for the whole
negative even grade mKdV subhierarchy, in particular for eqn. (\ref{19}).
Our solutions for eqn. (\ref{19}) does not appear in \cite{qiao}.
The introduction of the constant vacuum, $v_0$, showed that the simplest
1-soliton solution splits into three different classes, depending on the 
sign of the parameter $\gamma_1$ and its difference from $v_0$. 
The general form of the solutions agree with the
trivial vacuum ones, but its behavior is modified by the presence of 
the $v_0$ parameter. The 1, 2 and 3 soliton solutions were explicit checked
for eqn. (\ref{19}).  Moreover, the 1-soliton (\ref{1soliton})  with (\ref{data}) and $m=2$  was also verified to satisfy  eqn. (\ref{mkdvt-4}),
using symbolic computational methods.

\vskip 1cm

{\bf Acknowledgments.} 
\noindent {  
    We thank CNPq for support.
} \bigskip

\section{ Appendix - Matrix Elements}

Consider the vertex operator for $\hat{sl}(2)$,
\be
F\(\gamma, v_0\) = \sum_{n=-\infty}^{\infty}\(\gamma^2-v_0^2\)^{-n}
\bigg[ h^{(n)}+\frac{v_0-\gamma}{2\gamma}\delta_{n,0}\hc
+E_{\a}^{(n)}\(\gamma+v_0\)^{-1}-E_{-\a}^{(n+1)}\(\gamma-v_0\)^{-1}\bigg].
\ee

In the highest weight representation $\{|\lambda_0>, |\lambda_1>\}$ we
have the following action of $\hat{sl}(2)$  operators:
\br
E^{(0)}_{\alpha}|\lambda_a> & = & 0, \nonu \\
E^{(n)}_{\pm\alpha}|\lambda_a> & = & 0, \quad n > 0 \nonu \\
h^{(n)}|\lambda_a> & = & 0, \quad n > 0 \nonu \\
h^{(0)}|\lambda_a> & = & \delta_{a1}|\lambda_a> \nonu \\
\hat{c}|\lambda_a> & = & |\lambda_a>
\label{highest_states}
\er
where $a=0,1$. Using the adjoint relations $\(h^{(n)}\)^{\dagger}=h^{(-n)}, \;
\(E_{\alpha}^{(n)}\)^{\dagger} = E_{-\alpha}^{(-n)}$ and 
$\hat{c}^{\dagger}=\hat{c}$ we also know their actions on $<\lambda_a|$. 
From this, we have:
\br
<\l_0|F\(\gamma,v_0\)|\l_0> & = & \frac{\(v_0-\gamma\)}{2\gamma} 
\; \equiv \; c^{-}, 
\nonu \\
<\l_1|F\(\gamma,v_0\)|\l_1> & = & \frac{\(v_0+\gamma\)}{2\gamma} 
\; \equiv \; c^{+}.
\label{matrix_1}
\er

In order to calculate $<\l_a|F\(\gamma_1,v_0\)F\(\gamma_2,v_0\)|\l_a>$,
after distributing the products and keeping only non-trivial terms,
we make use of the commutator rules to change the order. 
The double sum simplifies to a single sum, which can then be
substituted for power series like 
$\sum_{n=0}^{\infty}x^n = 1/(1-x), \; \sum_{n=1}^{\infty}x^n = x/(1-x)$ and
$\sum_{n=1}^{\infty}nx^n = x/(1-x)^2$. The result is then given by:
\bdm
<\l_a|F\(\gamma_1,v_0\)F\(\gamma_2,v_0\)|\l_a>  \; = \; 
\delta_{a1} + 
\frac{2\(\gamma_1^2-v_0^2\)\(\gamma_2^2-v_0^2\)}{\(\gamma_1^2-\gamma_2^2\)^2}
+\frac{v_0-\gamma_1}{2\gamma_1}\delta_{a1} +
\edm
\bdm +\frac{v_0-\gamma_2}{2\gamma_2}\delta_{a1}
+\frac{\(\gamma_1-v_0\)\(\gamma_2-v_0\)}{4\gamma_1 \gamma_2}
-\frac{\(\gamma_1-v_0\)\(\gamma_2+v_0\)}{\gamma_1^2-\gamma_2^2}\delta_{a1} 
+\frac{\(\gamma_1+v_0\)\(\gamma_2-v_0\)}{\gamma_1^2-\gamma_2^2}\delta_{a1} - 
\edm
\be
-\(\gamma_1-v_0\)\(\gamma_2+v_0\)\frac{\gamma_2^2-v_0^2}{\(\gamma_1^2-\gamma_2^2\)^2}
-\(\gamma_1+v_0\)\(\gamma_2-v_0\)\frac{\gamma_1^2-v_0^2}{\(\gamma_1^2-\gamma_2^2\)^2}.
\label{v_square_all}
\ee
This expression can be further simplified to
\br
<\l_0|F\(\gamma_1,v_0\)F\(\gamma_2,v_0\)|\l_0> & = &
\frac{\(\gamma_1-v_0\)\(\gamma_2-v_0\)}{4\gamma_1\gamma_2}\(
\frac{\gamma_1-\gamma_2}{\gamma_1+\gamma_2}\)^2 \; = \; c_1^-c_2^-a_{12}, 
\nonu \\
<\l_1|F\(\gamma_1,v_0\)F\(\gamma_2,v_0\)|\l_1>  & = &
\frac{\(\gamma_1+v_0\)\(\gamma_2+v_0\)}{4\gamma_1\gamma_2}\(
\frac{\gamma_1-\gamma_2}{\gamma_1+\gamma_2}\)^2 \; = \;c_1^+c_2^+a_{12},
\nonu \\
\label{matrix_2}
\er
where $c^\pm_i = c^\pm\(\gamma_i\)$, see (\ref{matrix_1}), and we 
have defined:
\be
a_{ij} = \(\frac{\gamma_i-\gamma_j}{\gamma_i+\gamma_j}\)^2.
\ee
Note that (\ref{matrix_2}) $\rightarrow 0$
when $\gamma_2 \rightarrow \gamma_1$. This proves  the nilpotency property of 
the vertex operator when evaluated  within diagonal states $|\l_0>$ and  $|\l_1>$.

A more tedious  calculation  shows that:
\br
<\l_0|F\(\gamma_1,v_0\)F\(\gamma_2,v_0\)F\(\gamma_3,v_0\)|\l_0> & = &
c_1^-c_2^-c_3^-a_{12}a_{13}a_{23}, \nonu \\
<\l_1|F\(\gamma_1,v_0\)F\(\gamma_2,v_0\)F\(\gamma_3,v_0\)|\l_1> & = &
c_1^+c_2^+c_3^+a_{12}a_{13}a_{23}.
\er
In general, using Wick theorem, it is possible to show that:
\br
<\lambda_0| \prod_{i=1}^{n}F\(\gamma_i, v_0\) |\lambda_0> &=&
\prod_{i=1}^{n}c_i^- \prod_{i,j=1,\; i<j}^{n}a_{ij}, \nonu \\
<\lambda_1| \prod_{i=1}^{n}F\(\gamma_i, v_0\) |\lambda_1> &=&
\prod_{i=1}^{n}c_i^+ \prod_{i,j=1,\; i<j}^{n}a_{ij}.
\er



\end{document}